\documentstyle[prd,aps,12pt]{revtex}

\begin{document}

\draft
\title{Precision Theorectical Tests of Effective Lagrangians:\\
Pion Photoproduction from Threshold\\
Through the Delta Resonance\thanks{Invited talk, MAX-lab
Workshop on the Nuclear Physics Programme with Real Photons
below 200 MeV, Lund, Sweden, March 10-12, 1997. Presented by
N.C. Mukhopadhyay.}}
\author{ Nimai C. Mukhopadhyay and R. M. Davidson\\
Department of Physics, Applied Physics and Astronomy\\
Rensselaer Polytechnic Institute\\
Troy, NY 12180-3590, U.S.A.}

\maketitle

\begin{center}Abstract\end{center}

\begin{abstract}
We shall review here our current knowledge and new frontiers of
the pion photoproduction study, from threshold through the Delta
resonance region, emphasizing the hadron physics program, of
interest to a 250 MeV, high duty factor electron facility, such as
the upgraded MAX-lab at the Lund University, Sweden, where it would be
possible to produce tagged photons of high quality, with energy
below approximately 200 MeV.
\end{abstract}

\section{Introduction}

We are happy to be able to share with you our enthusiasm for physics
that can be done in the future with a high-quality real photon beam
below 200 MeV, such as that planned for the new MAX-lab. At the outset,
you may think that this is too low an energy region to be of much
interest to hadron physics, in particular, there being excellent
existing facilities in this energy region elsewhere. While it is
true that photon facilities at places like
Saskatoon, Mainz and others have had a go at this physics already,
much still remain to be done at {\it a high precision}. The advent and
success of chiral perturbation theory ($\chi$PT) \cite{ref1}
in explaining the pion photoproduction very near threshold have raised
our expectations of a QCD-based understanding of this region, where
already a solid understanding has existed on the basis of effective
Lagrangians motivated by chiral symmetry \cite{ref2}. In particular,
the roles of the matter fields, such as $\rho$ and $\omega$ mesons
and the $\Delta$(1232) resonance, which we have recently stressed
in our work \cite{ref2}, remain to be explored carefully, both
theoretically and experimentally. We believe future facilities like
the MAX-lab will have a crucial role in this. We also remain
concerned at the future of facilities like SAL at Saskatoon, Canada,
which is already doing excellent exploaration of this physics, as
is nicely represented in several contributions to this workshop. We
refer to these talks, as well as the theoretical review of the
$\chi$PT by J. Bijnens at this workshop for a nice background to
the physics issues of interest here.

In order to make our discussion a bit more general, we want to ask here
the question: What can we learn from a high-quality $E_{\gamma}$
$\le$ 200 MeV facility, in terms of precision theoretical tests of
effective Lagrangians? This means probing $\chi$PT, QCD and/or
whatever else you have for hadron structure.
The facilities where we can do this include Mainz,
SAL, MAX-Lab, TUNL, LEGS and others of this category. We shall also
assume that polarized photons and polarized targets will be eventually
available for experimental use. These tools add considerably to our
arsenals in order to probe relatively sensitive small amplitudes. 

A few short remarks on the history of ths subject, which is indeed a very
old one. The Delta resonance was discovered by Fermi, Steinberger and
their collaborators in the late forties to early fifties.
The first low-energy theorem (LET) was proposed by Kroll and Ruderman, and
the first dispersion theory of pion photoproduction by Chew {\it et al.}
(CGLN), all in the
fifties. The developments in the current algebra by Gell-Mann, Nambu,
Weinberg, Adler, Fubini and others in the sixties led to the LET's
as we know them. In the seventies, $\chi$PT got developed, thanks
largely due to
the efforts by Pagels, Weinberg, Leutwyler and Gasser. We have stressed
the roles of the $\Delta$, $\rho$, $\omega$ in the famous $E_{0+}$
amplitude for neutral pion production, in the eighties \cite{ref2}.
Then came the discovery of Mei{\ss}ner and collaborators \cite{ref1}
of the vital role of pion loops in the $E_{0+}$ amplitude, later
verfied by the experiments \cite{ref3}.

On the nuclear physics side, the dominance of the Gamow-Teller operator
in the charged pion photoproduction near threshold led in the early
seventies to selected nuclear excitation of the Gamow-Teller states
\cite{ref4} in these and related processes (electromagnetic excitation,
weak processes such as $\beta$ decay and $\mu$-capture). Finally, the
prominent roles of the Gamow-Teller excitations in the (p,n) and
(p,n) processes were found \cite{ref5}. Recently, even the excitation of
Gamow-Teller states in the neutrino reactions have become routine.

For the remainder of this paper, we shall cover the following
topics. In section II, we discuss what is new in the $\Delta$(1232)
electromagnetic excitation. In section III, we examine how the effective
Lagrangian approach (ELA), so successful at the Delta peak, does
near the pion photoproduction threshold, discussing the similarity
and differences of $\chi$PT and the simpler ELA's at low energy. In
section IV, we examine the critical energy region of interest to
the future MAX-Lab program. We conclude with a summary.

\section{What is new in the $\Delta$(1232) electromagnetic excitation?}

Thanks to recent experimental studies using polarized photons at the
Brookhaven LEGS \cite{ref6} and the more recent ones at the Mainz
Microtron \cite{ref7}, via the reaction $\vec{\gamma}p\rightarrow p
\pi^0$, we can infer \cite{ref8}, in the framework of the ELA, the
following values of the resonant $N \rightarrow \Delta$ helicity
amplitudes:
\begin{eqnarray}
A_{1/2} &=& -127.8 \pm 1.2 \nonumber \\
A_{3/2} &=& -252.4 \pm 1.3 \; ,
\end{eqnarray}
in the units of $10^{-3}$ GeV$^{-1/2}$. These come from the more
precise Mainz results. Translated into the resonant magnetic dipole
$(M1)$ and electric quadrupole $(E2)$ amplitudes, we get from (1),
the values, in the same units,
\begin{eqnarray}
M1 &=& 282.5 \pm 1.3 \nonumber \\
E2 &=& 9.00 \pm 0.66 \; ,
\end{eqnarray}
thereby yielding an $E2/M1$ ratio $(R_{EM})$
\begin{equation}
R_{EM}= -(3.19\pm 0.24)\% \; .
\end{equation}
Important points to note here are that the $M1$ amplitude is
substantially {\it larger} than the quark model (QM) estimates, around
200; the $E2$ is also much larger than the QM values \cite{ref9}. 
\begin{it}This
is a critical discrepancy in the QM, origins of which are as yet unclear.
\end{it}

These important amplitudes have been ``measured" on the lattice
for the first time. the results are \cite{ref10}:
\begin{equation}
M1=231\pm 41 \qquad \qquad R_{EM}=(3\pm 8)\% \; .
\end{equation}
The discrepancy of the $M1$ amplitude between (2) and (4) is there,
but not as severe as that with the QM. The $R_{EM}$ is too noisy to
be of decisive importance. Clearly, there is a lot of room for
improvements in the lattice ``measurement".

\section{Predictions for the effective Lagrangian approaches for
pion photoproduction near threshold}

The chiral symmetry constraints put the conventional ELA with
considerable predictive power in describing charged pion
photoproduction near threshold. However, there is a large correction
in the $E_{0+}$ amplitude for the neutral pion photoproduction.
Thus, for the reaction $\gamma p \rightarrow p \pi^0$, the old
tree-level results yield, for s- and u- channel exchange \cite{ref11},
\begin{equation}
E_{0+} = -2.23 \; ,
\end{equation}
in the usual units of $10^{-3}/M_{\pi^+}$. We have shown \cite{ref2}
that the t-channel vector meson exchanges and the $\Delta$ exchange
change this number to
\begin{equation}
E_{0+} \approx -1.94 \; .
\end{equation}
The $\chi$PT result, first established by Bernard {\it et al.}
\cite{ref1}, gives a sizable one-pion loop correction to yield
\begin{equation}
E_{0+} \approx -1.00 \; .
\end{equation}

The current situation for the $\pi^0$ threshold multipoles is
summarized in Table I. Clearly, the $\chi$PT results are the closest
to the experiment in the $E_{0+}$ case, while our theoretical
tree-level results fare as well in the cases of the multipoles
$P_1$ and $P_{23}$. The $\chi$PT results for the s-wave multipole
still suffer from considerable uncertainty. For the p-waves, \begin{it}
clearly
the matter fields play important roles\end{it} (Table II).

\section{Similarities and differences between $\chi$PT and simpler
effective Lagrangians}

Table I already gives us a strong indication of the similarities
and differences between predictions of $\chi$PT and simpler ELA's for
$\pi^0$ near threshold. As matter fields become stronger in the p-wave
multipoles, the disagreement between $\chi$PT and the ELA of the simpler
sort becomes more acute. Thus, for $E_{\gamma}$ $\ge$ 160 MeV, these
differences begin to stand out. Thus, \begin{it}
the energy region $E_{\gamma}$
$\sim$ 170 to 200 MeV is quite crucial to sort out this difference
between the two classes of theories which begin to show their
dynamical difference.\end{it}
Experiments are needed to precisely map out
this region and \begin{it}the MAX-Lab can be of vital help in this.
\end{it}

\section{Conclusions}

{}From this survey, the following points can be stressed:

(1) The $E_{0+}$ multipole, probed by the $\pi^0$ photoproduction
near threshold, is an important test of $\chi$PT, where the
conventional ELA's at the tree level do emphasize some role of the
matter fields.

(2) The multipoles like $M_{1+}$ favor the conventional ELA's
which provide efficient treatments of the $\Delta$ resonance.
The $\chi$PT has some work to do to clean up this physics.

(3) The transition region between these two classes of theories
for $E_{\gamma}$ is from 170 to 200 MeV.
Considerable theoretical and experimantal works are needed here to make
the $\chi$PT tests a precise theoretical enterprise.

Thus, the MAX-Lab and labs like that have an important QCD
mission in this energy domain.

Let us max the MAX! This is one of the best ways to celebrate at Lund
the memory of the late Professor Janne Rydberg in the next century.

\section{Acknowledgments}

One of us (NCM) is grateful to Dr. Bent Schr{\"o}der for a very kind
invitation and gracious hospitality which made his trip to Sweden
possible. We also thank our undergraduate colleague, Michael Pierce,
for numerical help.
Our research is supported by the U.S. Dept.~ of Energy.

\vspace{24pt}

\begin{table}[top]
\caption{Threshold $\pi^0$ multipoles in the $\chi$PT
\protect\cite{ref1}, in the simpler effective Lagrangian approach
(DMW) at the tree level \protect\cite{ref2} and in the recent
experiment of Fuchs {\it et al.} \protect\cite{ref3}. The multipoles
are in the usual units.}
\begin{center}
\begin{tabular}{|l|c|c|c|} \hline
  & $\chi$PT & DMW & Expt.  \\
$E_{0+}$ & -1 to -1.5 & -2.15 $\pm$ 0.23 & -1.31 $\pm$ 0.08   \\
$P_1$ & 10.3 & 11.28 $\pm$ 0.14 & 10.02 $\pm$ 0.15  \\
$P_{23}$ & 11.25 & 11.85 $\pm$ 0.20 & 11.44 $\pm$ 0.08 \\ \hline
\end{tabular}
\end{center}
\end{table}

\begin{table}[bottom]
\caption{Importance of matter fields in controlling the $\pi^0$
photoproduction multipoles at threshold. The contributions
enumerated are evaluated from the fits of Davidson {\it et al.}
(DMW) \protect\cite{ref2}.}
\begin{center}
\begin{tabular}{|c|c|c|c|c|}\hline
 & PV Born & Vector Mesons & Delta & Delta  \\
  &  &  & $g_1$ coupling & $g_2$ coupling \\
$E_{0+}$ & -2.47 & +0.08 & -0.05 & +0.40 \\
$M_{1-}$ & -6.57 & +0.89 & +2.03 & +0.19 \\
$E_{1+}$ & +0.04 & -0.01 & -0.15 & +0.10 \\
$M_{1+}$ & +3.39 & +0.74 & +3.96 & -0.14 \\ \hline
\end{tabular}
\end{center}
\end{table}

\end{document}